\documentclass[12pt]{article} 

\usepackage{epsf} 

\newcommand{\beq}{\begin{equation}} 
\newcommand{\eeq}{\end{equation}} 
\newcommand{\beqa}{\begin{eqnarray}} 
\newcommand{\eeqa}{\end{eqnarray}}

\newcommand{\no}{\nonumber}

\newcommand{\vs}{\vspace{-0.25cm}} 
\newcommand{\dfrac}{\displaystyle \frac} 
\newcommand{\mc}{\multicolumn{2}{c||}} 
\newcommand{\mcc}{\multicolumn{3}{c||}} 

\begin{document} 
    
\hfill {\small FZJ-IKP(TH)-2000-25}

\bigskip\bigskip\bigskip 
 
\begin{center} 

{{\Large\bf Baryon form factors in \\[0.3em] 
   chiral perturbation theory\footnote{Work supported in part
   by funds provided by the Graduiertenkolleg ``Die Erforschung subnuklearer
   Strukturen der Materie'' at Bonn University
   and by the ``Studienstiftung des deutschen Volkes''.}
}} 
 
\end{center} 
 
\vspace{.2in} 
 
\begin{center} 
{\large  Bastian Kubis\footnote{E-mail: b.kubis@fz-juelich.de},
Ulf-G. Mei{\ss}ner\footnote{E-mail: Ulf-G.Meissner@fz-juelich.de}}

\bigskip 
 
\bigskip 
 
{\it Forschungszentrum J\"ulich,  
Institut f\"ur Kernphysik (Theorie)\\  
D--52425 J\"ulich, Germany}

\end{center} 
 
\vspace{.7in} 
 
\thispagestyle{empty}  
 
\begin{abstract} 
\noindent 
We analyze the electromagnetic form factors of the ground state baryon octet 
to fourth order in relativistic baryon chiral perturbation theory.  
Predictions for the $\Sigma^-$ charge radius and the
$\Lambda$--$\Sigma^0$ transition moment are found to be in excellent
agreement with the available experimental information.
Furthermore, the convergence behavior of the hyperon charge radii
is shown to be more than satisfactory.
\end{abstract} 
 
\vspace{1.3in} 
 
 
\centerline{Keywords:  
{\it Baryon form factors}, {\it chiral perturbation theory}} 
 
\vfill

\newpage 
 
\section{Introduction} 
\label{sec:intro} 
\def\theequation{\arabic{section}.\arabic{equation}} 
\setcounter{equation}{0} 

Hadrons are composite objects, characterized by certain
probe--dependent sizes. Their structure can be investigated by use of
electron scattering (or the inverse process). The electromagnetic structure
of the proton and the neutron has been investigated over decades,
the present status of the data is e.g.\ discussed in~\cite{petn99}. In the
non--perturbative low--energy region of QCD,
baryon chiral perturbation theory can be used
to calculate these form factors. In a recent paper~\cite{irff} we have
shown that relativistic baryon chiral perturbation theory (employing the
so--called infrared regularization of~\cite{Becher})
supplemented by
explicit vector meson contributions allows for a fairly precise description
of these fundamental quantities for photon virtualities up to $Q^2 \simeq 0.4$~GeV$^2$.
The extension of these considerations to the three--flavor case is interesting
for various reasons. First, the charge radius of the $\Sigma^-$ has recently
been measured~\cite{wa89,selex} and thus gives a first glimpse of an electric hyperon form factor.
Second, chiral SU(3) can be subject to large kaon/eta loop corrections, 
and the form factors offer another window
to study the corresponding convergence properties. 
They might thus indicate whether or not the strange quark can be considered light 
and lead to a better understanding of SU(3) flavor breaking. 
Since the chiral expansion of the form factors is well under control 
in the two--flavor case, one can expect
to encounter a reasonably well--behaved series also in the presence of the
strange quark. This expectation is borne out by the results presented in
this paper. 
Third, one can also address some questions concerning 
strangeness in the nucleon, more precisely, the role
of kaon loops which in simple models let one expect sizeable contributions of strange
operators.
Fourth, a knowledge of certain hyperon form factors is mandatory to gain an understanding
of kaon photo-- and electroproduction off nucleons and light nuclei
as measured at ELSA and TJNAF. We will come back
to most of these  topics in the present manuscript.  In addition, these form
factors have already been calculated in the so--called heavy--baryon
approach~\cite{KHM,puglia2}, which is a particular limit of the regularization
procedure employed here. A direct comparison with the results of that
approach can shed further light on the dynamics underlying the
non--perturbative baryon structure, in particular the role of recoil
corrections. 
As a final by--product, we can also readdress the issue
of the convergence of the chiral expansion for the magnetic moments,
which is much discussed in the recent literature~\cite{jlms,summ,DH,KM,puglia1}.

\medskip \noindent
The manuscript is organized as follows. In section~\ref{sec:ff} we briefly
define the baryon form factors and the corresponding electromagnetic radii.
The formalism to obtain the one--loop representation of the form factors is given in
section~\ref{sec:1loop}. We heavily borrow from~\cite{irff} and omit
all lengthy formulae. The results are presented and discussed in
section~\ref{sec:results}. Section~\ref{sec:sum} contains a short summary
and outlook.

\section{Baryon form factors} 
\label{sec:ff} 
\def\theequation{\arabic{section}.\arabic{equation}} 
\setcounter{equation}{0} 
 
The structure of ground state octet baryons (denoted by `$B$')
as probed by virtual photons is parameterized in terms of two form factors each, 
\beq 
\langle B(p')\, | \, {\cal J}_\mu \,  | \, B(p)\rangle  
= e \,  \bar{u}(p') \, \biggl\{  \gamma_\mu F_1^B (t) 
+ \frac{i \sigma_{\mu \nu} q^\nu}{2 m_B} F_2^B (t) \biggr\}  
\,  u(p) \,, \quad B=p,n,\Lambda,\Sigma^\pm,\Sigma^0,\Xi^-,\Xi^0 \,, 
\label{FFdef}
\eeq 
with $t = q_\mu q^\mu = (p'-p)^2$ the invariant momentum  
transfer squared, ${\cal J}_\mu$  the quark vector current,
${\cal J}_\mu = \bar{q} {\cal Q} \gamma_\mu q$ (${\cal Q}$ is the quark
charge matrix and $q^T=(u,d,s)$),
and $m_B$ the respective baryon mass.
In electron scattering $t$ is negative and it is often convenient  
to define the positive 
quantity $Q^2 = -t > 0$. $F_1$ and $F_2$ are called the Dirac and the Pauli 
form factor, respectively, with the normalizations $F_1^B (0) =Q_B$, 
$F_2^B (0) =\kappa_B$. Here $\kappa_B$ denotes the anomalous magnetic moment. 
Eq.~(\ref{FFdef}) has to be generalized for the 
$\Lambda$--$\Sigma^0$ transition form factors which are 
defined according to
\beq 
\langle \Sigma^0(p')\, | \, {\cal J}_\mu \,  | \, \Lambda(p)\rangle  
= e \,  \bar{u}(p') \, \biggl\{  
\biggl( \gamma_\mu - \frac{m_{\Sigma^0}\!-\! m_\Lambda}{t} \, q_\mu \biggr)
F_1^{\Lambda\Sigma^0}(t) 
+ \frac{i \sigma_{\mu \nu} q^\nu}{m_\Lambda\!+\!m_{\Sigma^0}} F_2^{\Lambda\Sigma^0} (t) \biggr\}  
\,  u(p) 
\label{FFLS0def}
\eeq 
(see also~\cite{LWD}).
The form of the generalized Lorentz structure accompanying $F_1^{\Lambda\Sigma^0}(t)$
is required by current conservation (which becomes obvious by contracting
eq.~(\ref{FFLS0def}) with $q^\mu$ and applying the Dirac equation). 
Different definitions which all reduce to eq.~(\ref{FFdef}) for $m_{\Sigma^0}\rightarrow m_\Lambda$
are possible, however, the one given here is preferable because it still requires
the normalization $F_1^{\Lambda\Sigma^0}(0)=0$.
One also uses the electric and magnetic Sachs form factors, 
\beq 
G_E (t) = F_1 (t) + \dfrac{t}{4m_B^2} F_2 (t) \, ,
\quad G_M (t) = F_1 (t) + F_2 (t) \, ,
\label{sachsdef}
\eeq
which are the quantities we will consider in the following.
The slope of the form factors at $t=0$ is conventionally expressed in 
terms of a radius $\langle r^2\rangle^{1/2}$, 
\beq 
F(t) = F(0) \, \biggl( 1 + \dfrac{1}{6}\langle r^2\rangle \, t + \ldots \biggr)
\eeq 
($F$ being a genuine symbol for any of the four electromagnetic baryon form factors),
and the mean square radius of this charge distribution is given by 
\beq 
\langle r^2\rangle = 4 \pi\, \int_0^\infty dr \,  r^2 \rho(r) = 
\dfrac{6}{F(0)} \dfrac{dF(t)}{dt} \biggr|_{t = 0} ~. 
\label{defrad}   
\eeq 
Eq.~(\ref{defrad}) can be used for all form factors except for the electric ones
of the neutral baryons which vanish at $t=0$. In these cases, one simply drops the 
normalization factor $1/F(0)$ and defines e.g.\ the neutron charge radius 
via 
\beq 
\langle (r_E^n)^2 \rangle \,  = 6 \, \dfrac{dG_E^n(t)}{dt} 
\biggr|_{t = 0} ~.
\eeq 
 

\section{Formalism}
\label{sec:1loop}
\def\theequation{\arabic{section}.\arabic{equation}} 
\setcounter{equation}{0} 

In this section, we spell out the details necessary to extend the SU(2)
calculation of~\cite{irff} to the three--flavor case. We work in relativistic
baryon chiral perturbation theory, employing infrared regularization (IR).
For details on this procedure, we refer to~\cite{Becher,irff}.
Although the presence of different pseudo--Goldstone bosons with unequal masses
entails more general loop functions than those encountered in~\cite{irff}, 
we refrain from tabulating these here. 
They will become available in~\cite{futwon}. We only spell out the various terms
of the effective chiral Lagrangian underlying the calculation. 
Again we do not give the final formulae for the form factors since these are rather
lengthy, but instead refer to~\cite{futwon}.


\subsection{Effective Lagrangian} 
\label{sec:lagr}

\medskip\noindent 
The chiral effective Goldstone boson Lagrangian is given by
\beq
{\cal L}_{\phi\phi}^{(2)} = \frac{F^2}{4}\langle u_\mu u^\mu +\chi_+ \rangle ~,
\eeq
where the octet of Goldstone boson fields is collected in the SU(3) valued matrix 
$U(x) = u^2(x)$, and the chiral vielbein is related to $u$ via
$u_\mu = i\{u^\dagger , \nabla_\mu u\}$.
$\nabla_\mu$ is the covariant derivative acting on the pion fields including
external vector ($v_\mu$) and axial ($a_\mu$) sources,
$\nabla_\mu U= \partial_\mu U -i\bigl(v_\mu+a_\mu\bigr)U+iU\bigl(v_\mu-a_\mu\bigr)$. 
The mass term is included in the field $\chi_+$ via the definitions $\chi = 2B(s+i\,p)$
and $\chi_+=u^\dagger \chi u^\dagger + u \chi^\dagger u$, with $s$ and $p$ being
scalar and pseudoscalar sources, respectively. $s$~includes
the quark mass matrix, $s={\rm diag}(m_u,m_d,m_s)+\ldots$~. We will work in
the isospin limit, $m_u=m_d=\hat{m}$. 
$B=|\langle 0|\bar{q}q|0\rangle |/F^2$ measures 
the strength of the symmetry violation, and we assume the standard
scenario, $B \gg F$.
Furthermore,  $\langle \ldots \rangle$ denotes the trace in flavor space.
The mass term leads to the well--known lowest--order (isospin symmetric) mass formulae
for pions, kaons, and the eta, which enter the description of the baryon form factors
via loop contributions. We would like to point out that 
the mass differences between the Goldstone bosons 
yield the leading--order SU(3) breaking effect for these form factors.
For numerical evaluation, we will use $M_\pi=139.57$~MeV, $M_K=493.68$~MeV
(the \emph{charged} pion and kaon masses), and $M_\eta=547.45$~MeV.
Furthermore, it is legitimate to differentiate between different
decay constants $F_\pi$, $F_K$,
$F_\eta$ in the treatment of the chiral loops as these differences are of higher order.
We will use $F_\pi=92.4$~MeV, $F_K/F_\pi=1.22$, $F_\eta/F_\pi=1.3$
(see~\cite{GLNPB} and, for a more recent determination
of $F_K/F_\pi$, \cite{FKS}). The main motivation for not
using a common decay constant
is the comparison to the SU(2) results for the proton and neutron form factors, where
we do not want to suggest an SU(3) effect which is only due to a numerically different
treatment of the pion loops.

\medskip \noindent
The meson--baryon Lagrangian at leading order reads 
\beq 
{\cal L}_{\phi B}^{(1)} = \langle
\bar{B} \left( iD\!\!\!\!/ - m \right) B \rangle
+ \frac{D/F}{2} \, \langle \bar{B} \gamma^\mu \gamma_5
\left( u_\mu ,B\right)_\pm \rangle  ~, 
\eeq 
where the matrix--valued field $B$ collects the ground state octet baryons
and $D$ and $F$ are the axial vector coupling constants.\footnote{Here and in
  what follows, we employ a compact notation: the $D$--type coupling refers
  to the anticommutator and the $F$--type coupling to the commutator.}
For these, we will use
the values $D=0.80$, $F=0.46$ extracted from hyperon decays~\cite{ratcliffe},
which obey the SU(2) constraint $D+F=g_A=1.26$. Here, $m$ denotes the
average baryon mass in the chiral limit.
To this order, the photon field only couples to the charge of the baryon.
It resides in the chiral covariant derivative,
$D_\mu B= \partial_\mu B + [\Gamma_\mu,B]$,
with the chiral connection given by
$\Gamma_\mu = \frac{1}{2}\bigl[u^\dagger, \partial_\mu u\bigr]
-\frac{i}{2}u^\dagger (v_\mu+a_\mu)u -\frac{i}{2}u(v_\mu-a_\mu)u^\dagger$.

\medskip \noindent 
Coupling constants from the second--order meson--baryon Lagrangian are needed both
at tree level and in one--loop graphs. The following terms are required in our calculation:
\beqa 
{\cal L}_{\phi B}^{(2)} &=&
b_{D/F} \langle \bar{B} ( \chi_+ , B )_\pm \rangle
+ \frac{b_6^{D/F}}{8m} \langle \bar{B} \sigma^{\mu\nu}
(F_{\mu\nu}^+ , B )_\pm \rangle \no\\
&& + \frac{i}{2}\sigma^{\mu\nu} \biggl\{ b_9 \langle \bar{B} u_\mu \rangle \langle u_\nu B \rangle
   + b_{10/11} \langle \bar{B} ( [u_\mu,u_\nu], B)_\pm \rangle \biggr\} ~.
\label{Lagr2}
\eeqa
Here, $F^+_{\mu\nu}=u^\dagger F_{\mu\nu} u + u F_{\mu\nu} u^\dagger$,
and $F_{\mu\nu} = \partial_\mu A_\nu - \partial_\nu A_\mu$
is the conventional photon field strength tensor.
The $b_i$  are the so--called low--energy constants (LECs)
which encode information about the more massive states not contained in the 
effective field theory or other short--distance effects.
These parameters have to be pinned down by using some data.
In principle there is also a term 
$\sim b_0 \, \langle \bar{B} B \rangle \langle \chi_+ \rangle$
which amounts to a quark mass renormalization of the
common octet mass $m$. This, however, cannot be disentangled from $m$ without further
information (like the pion--nucleon $\sigma$ term), and it is sufficient 
for our purpose to absorb this term in $m$.
The couplings $b_{D/F}$ yield the leading SU(3) breaking effects in the baryon masses.
Again, these affect the form factors via various loop contributions.
A best fit to the octet masses results in $m_N=0.942$~GeV,
$m_\Lambda=1.111$~GeV,  $m_\Sigma=1.192$~GeV,
$m_\Xi=1.321$~GeV (for $m=1.192$~GeV, $b_D=0.060$~GeV$^{-1}$,
$b_F=-0.190$~GeV$^{-1}$), compared to
the experimental values $m_N=0.939$~GeV, $m_\Lambda=1.116$~GeV, $m_\Sigma=1.193$~GeV,
$m_\Xi=1.318$~GeV (where the average masses within the respective
isospin multiplets have been taken).
We consider this accurate enough to put all baryon
masses to their experimental values in the numerical evaluation.
In any third--order calculation, however, no mass splitting is present,
hence the baryon mass parameter will always be put 
to the average baryon mass $\bar{m}=1.151$~GeV in this case.
It has been discussed in detail in
\cite{Becher,irff} how to fix the omnipresent mass scale $\lambda$
which has to be introduced in loop diagrams
treated in dimensional regularization. 
It was argued that in an SU(2) calculation,
the nucleon mass serves as a natural mass scale. 
Here we consider it natural to set $\lambda=\bar{m}$.
The LECs $b_6^{D/F}$ parameterize the leading magnetic photon couplings to
the baryons and will be fitted to the magnetic moments.
Finally, the LECs $b_{9/10/11}$ accompany second--order couplings of the
various pseudo--Goldstone bosons
to baryons and enter the form factors via (tadpole) loop contributions.
Their values have been estimated
based on the resonance saturation hypothesis in~\cite{summ}, however,
the results used there do not
satisfy the SU(2) constraint $2(b_{10}+b_{11})=c_4 \approx 3.4$~GeV$^{-1}$.
The latter value is well--established, 
consistently determined from resonance saturation~\cite{BKMpin}
and fits to pion--nucleon scattering~\cite{FMS,Paul}. 
The discrepancy between both determinations can be traced back
to different treatments
of the $\Delta$ contribution in~\cite{summ} and~\cite{BKMpin}.
Adjusting this, we will use the values
$b_9=1.36$~GeV$^{-1}$, $b_{10}=1.24$~GeV$^{-1}$, $b_{11}=0.46$~GeV$^{-1}$.
As we would like to specify the uncertainty of our predictions based
on these estimates later on, we attribute some errors to these LECs. 
As an indication we regard the change of a LEC when fitting
it within chiral amplitudes of different orders.
This change is about 1.0~GeV$^{-1}$ for $c_4$ when going from second to third
order in $\pi N$ scattering (see~\cite{pipiN} for a second order fit).
We assume that this change should be a factor of 2 smaller when
proceeding from third to fourth order such that,
due to the different normalization of the SU(3) couplings, we
set $\Delta b_{9/10/11}=0.25$~GeV$^{-1}$.

\medskip \noindent
The only terms needed from the third order Lagrangian are
those entering the electric (charge) radii
of the baryons,
\beqa
{\cal L}_{\phi B}^{(3)} &=& \frac{i d_{101/102}}{2m} \biggl\{ \langle \bar{B}
\Bigl( [D^\mu,F_{\mu\nu}^+],[D^\nu,B] \Bigr)_\mp \rangle + {\rm h.c.} \biggr\} ~. 
\label{Lagr3}
\eeqa 
$d_{101},\,d_{102}$ have to be fitted to the charge radii of proton and neutron.

\medskip \noindent 
At fourth order, two types of coupling constants appear which are
of relevance for our calculation:
two couplings entering the magnetic radii, and seven couplings proportional
to a quark mass insertion
contributing to the magnetic moments,
\beqa 
{\cal L}_{\phi B}^{(4)} &=&
  \frac{\alpha_{1/2}}{8} \langle \bar{B} \sigma^{\mu\nu}
\Bigl(  [F_{\mu\nu}^+,B ],\chi_+ \Bigr)_\mp \rangle
+ \frac{\alpha_{3/4}}{8} \langle \bar{B} \sigma^{\mu\nu}
\Bigl( \{F_{\mu\nu}^+,B\},\chi_+ \Bigr)_\mp \rangle \no\\
&&+ \frac{\beta_1}{8} \langle \bar{B} \sigma^{\mu\nu} B\rangle
\langle \chi_+ F_{\mu\nu}^+ \rangle
+ \frac{\tilde{b}_6^{D/F}}{8} \langle \chi_+ \rangle
\langle \bar{B} \sigma^{\mu\nu} (F_{\mu\nu}^+ , B )_\pm \rangle \no\\
&&- \frac{\eta_{1/2}}{2} \langle \bar{B} \sigma^{\mu\nu}
\Bigl( [D_\lambda,[D^\lambda,F_{\mu\nu}^+]], B\Bigr)_\mp \rangle
~.  \label{Lagr4}
\eeqa 
As indicated by the notation, the terms $\sim \tilde{b}_6^{D/F}$ only amount
to a quark mass renormalization of the leading magnetic couplings
and will therefore be absorbed into $b_6^{D/F}$.
The LECs $\alpha_{1-4}$, $\beta_1$, however, incorporate explicit
breaking of SU(3) symmetry
in the magnetic moments and will be fitted to the octet moments.
$\eta_{1/2}$ will be adjusted to the magnetic radii of proton and neutron.

\medskip \noindent
While there are, all in all, quite some LECs to be fitted in order to describe
all electromagnetic form factors, it is important to point out that all these
LECs are, on general grounds, expected to be of order 1
(with the appropriate mass dimensions in powers of GeV:
as the baryonic scale is of the order of 1~GeV, it is not necessary to
normalize the LECs appropriately).\footnote{We would like to point out that
we have adopted the natural normalization for the SU(3) breaking
terms in the magnetic moments (by using the field $\chi_+$ proportional to the
quark mass matrix as defined above instead of a spurion ${\rm diag}(0,0,1))$
which is different from e.g.~\cite{jlms,summ}. Numerically, however,
this difference amounts to a factor of $4M_K^2\approx 0.975$~GeV$^2$, hence
apart from a different mass dimension, this does not make any significant difference.}

\subsection{Chiral expansion of the baryon form factors} 
 
The chiral expansion of a form factor $F$  consists of two contributions,  
tree and loop graphs. 
The tree graphs comprise the lowest--order diagram with fixed coupling (the baryon charge) 
as well as counterterms from the second--, third--, and fourth--order Lagrangians. 
As one--loop graphs we have both those with just lowest--order couplings 
and those with exactly one insertion from  ${\cal L}_{\phi B}^{(2)}$.  
The pertinent tree and loop graphs are depicted in fig.~\ref{fig:diag} 
(we have not shown the diagrams leading to wave function renormalization). 
We refrain from giving the explicit expressions here but we mention that
in the limit of a heavy strange quark, we recover the SU(2) results of
\cite{irff}. Also, the heavy--baryon (HB) results of~\cite{KHM} can be
obtained straightforwardly as detailed in~\cite{irff}.

\section{Results and discussion}
\label{sec:results}
\def\theequation{\arabic{section}.\arabic{equation}} 
\setcounter{equation}{0} 

We first discuss the issues concerning the magnetic moments and
the electromagnetic radii in some detail. 
This can be done within the `pure' chiral expansion.
Then we turn to the full momentum dependence of the
various form factors for photon virtualities up to $Q^2 = 0.3$~GeV$^2$.
To this end we have to include vector mesons as active degrees of freedom in a
chirally symmetric manner.

\subsection{Magnetic moments} 
\label{sec:magnmom} 

The issue of convergence of the magnetic moments in chiral perturbation theory (ChPT)
has been discussed amply in the literature, with or without
inclusion of the decuplet \cite{jlms,summ,DH,KM,puglia1}.
We wish to compare the convergence behavior in the heavy--baryon and the
infrared regularization scheme.
Table~\ref{tab:magnmom} shows the best fits to the various orders.
The strategy is always to fit the seven experimentally measured static moments
($\mu_{\Sigma^0}$ has not been measured so far, in the theoretical predictions
it is always given by $\mu_{\Sigma^0}=(\mu_{\Sigma^+}+\mu_{\Sigma^-})/2$
according to isospin symmetry)
and to predict the transition moment $\mu_{\Lambda\Sigma^0}$.
At second and third order, only the two leading--order couplings $b_6^{D/F}$ are free
parameters, which results in best fits of varying quality,
whereas at fourth order the additional five couplings 
allow for an exact fit of all seven static moments.
At second/third order, we have performed an unweighted fit, hence simply
minimizing $\chi^2=\sum(\mu_{\rm th}-\mu_{\rm exp})^2$.
The $\chi^2$ values given in table~\ref{tab:magnmom} thus only serve to indicate
the relative quality of the fits.

\begin{table}[ht] 
\begin{center} 
\renewcommand{\arraystretch}{1.2} 
\begin{tabular}{|c||c||c|c||c|c|c||c|} 
\hline
& & \mc{HB} & \mcc{IR} & \\
\hline 
& ${\cal O}(q^2)$ & ${\cal O}(q^3)$ & ${\cal O}(q^4)$ &
                    ${\cal O}(q^3)$ & ${\cal O}(q^3)^*$ &
                    ${\cal O}(q^4)$ & exp. \\ 
\hline 
$p$               &    2.56 &    2.97 &    2.793        &    2.61 &    2.20 &    2.793        & $ 2.793\pm0.000$ \\
$n$               & $-$1.60 & $-$2.53 & $-$1.913        & $-$1.69 & $-$2.59 & $-$1.913        & $-1.913\pm0.000$ \\
$\Lambda$         & $-$0.80 & $-$0.45 & $-$0.613        & $-$0.76 & $-$0.65 & $-$0.613        & $-0.613\pm0.004$ \\
$\Sigma^+$        &    2.56 &    2.21 &    2.458        &    2.53 &    2.41 &    2.458        & $ 2.458\pm0.010$ \\
$\Sigma^0$        &    0.80 &    0.45 &    0.649        &    0.76 &    0.65 &    0.649        &    ---           \\
$\Sigma^-$        & $-$0.97 & $-$1.32 & $-$1.160        & $-$1.00 & $-$1.12 & $-$1.160        & $-1.160\pm0.025$ \\
$\Xi^0$           & $-$1.60 & $-$0.78 & $-$1.250        & $-$1.51 & $-$1.20 & $-$1.250        & $-1.250\pm0.014$ \\
$\Xi^-$           & $-$0.97 & $-$0.56 & $-$0.651        & $-$0.93 & $-$1.33 & $-$0.651        & $-0.651\pm0.003$ \\
$\Lambda\Sigma^0$ &    1.38 &    1.65 &   $1.46\pm0.01$ &    1.41 &    1.81 &   $1.61\pm0.01$ & $\pm1.61\pm0.08$ \\
\hline
$b^D_6$           &    2.40 &    5.27 &   $4.56\pm0.24$ &    3.65 &    5.18 &   $4.21\pm0.20$ & \\
$b^F_6$           &    0.77 &    2.92 &   $1.65\pm0.19$ &    1.73 &    0.56 &   $1.64\pm0.18$ & \\
$\alpha_1$        &    ---  &    ---  &  $-1.00\pm0.26$ &    ---  &    ---  &   $0.32\pm0.28$ &\\
$\alpha_2$        &    ---  &    ---  &   $1.35\pm0.29$ &    ---  &    ---  &  $-0.08\pm0.18$ &\\
$\alpha_3$        &    ---  &    ---  &  $-0.85\pm0.30$ &    ---  &    ---  &   $2.14\pm0.28$ &\\
$\alpha_4$        &    ---  &    ---  &   $0.95\pm0.22$ &    ---  &    ---  &   $0.05\pm0.19$ &\\
$\beta_1$         &    ---  &    ---  &  $-2.46\pm0.33$ &    ---  &    ---  &  $-3.39\pm0.34$ &\\
\hline
$\chi^2$          &    0.46 &    0.76 &    0.00         &    0.28 &    1.28 &    0.00         & \\
\hline
\end{tabular} 
\renewcommand{\arraystretch}{1.0} 
\caption{
Analysis of the magnetic moments (in units of nuclear magnetons (n.m.)) to different chiral orders.
The various best--fit values for the leading order magnetic couplings
$b^{D/F}_6$ and the SU(3) breaking couplings $\alpha_{1-4}$, $\beta_1$
are given as well. $b^{D/F}_6$ are dimensionless, $\alpha_{1-4}$, $\beta_1$
are given in units of GeV$^{-3}$.
Errors for fourth--order results display the uncertainty due to $\Delta b_{9/10/11}$.
For the definition of $\chi^2$, see text.} 
\label{tab:magnmom} 
\end{center} 
\end{table}  

\begin{figure}[htb]
\centerline{
\epsfysize=2.8in
\epsffile{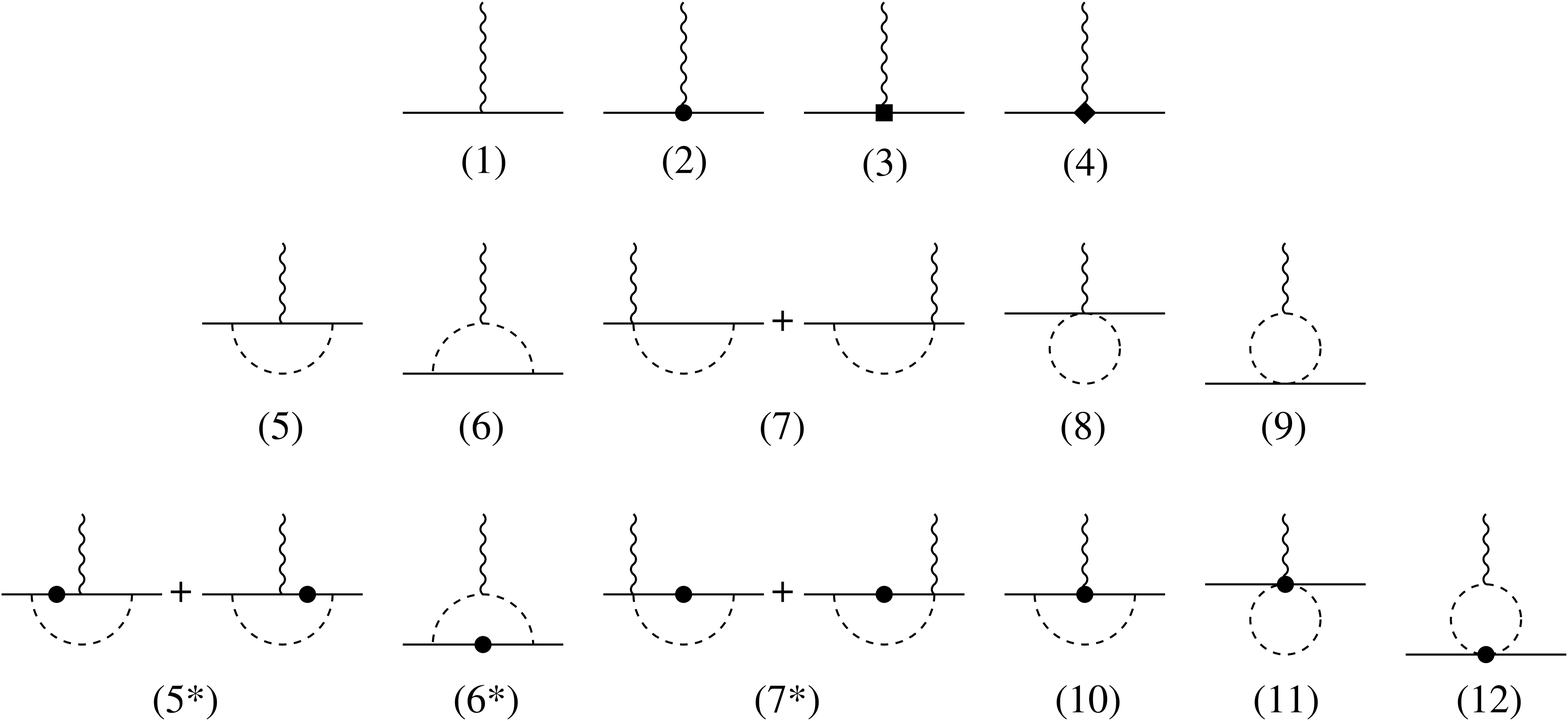}
}
\vskip 2.5cm
\caption{Feynman diagrams contributing to the electromagnetic form factors
up to fourth order. Solid, dashed, and wiggly lines refer to baryons,
Goldstone bosons, and the vector source, respectively. Vertices denoted by a heavy
dot/square/diamond refer to insertions from the
second/third/fourth order chiral Lagrangian, respectively.
Diagrams contributing via wave function renormalization only are not shown.
\label{fig:diag}
}
\end{figure}

\medskip \noindent
It has frequently been noted before that the inclusion of leading loop corrections
in the magnetic moments tends to \emph{worsen} the leading--order results.
This is seen here in the third--order heavy--baryon results. Within the
infrared regularization scheme, however, it is even disputable which
contributions to count as third order: the leading contributions stem from
diagram (6) in fig.~\ref{fig:diag}; summing up only the $1/m$--corrections
to this graph yields the fit in the column denoted by `${\cal O}(q^3)$' and
shows an improvement not only over the heavy--baryon result, but also over the leading
order fit. However, defining any one--loop diagram with no higher--order insertions
to be of third order, one also has to include diagram (5) in fig.~\ref{fig:diag}
(which only contributes at fourth order in strict chiral power counting). These
contributions are large and worsen the fit even over the third--order heavy--baryon one,
see the column denoted by `${\cal O}(q^3)^*$'.
Especially the magnetic moments of proton, neutron, and $\Xi^-$ are not described
to any acceptable accuracy in this case. We note furthermore that the fitted values for the
leading--order couplings vary considerably among the different fits.
Regarding these as indicators for convergence, we again find that
the `${\cal O}(q^3)$' result is much closer to the fourth--order fit than
the `${\cal O}(q^3)^*$' one.

\medskip \noindent
At fourth order, we can compare the two predictions
for the transition moment $\mu_{\Lambda\Sigma^0}$.
In both cases, we have indicated the uncertainty of these predictions due to the
estimated uncertainties of the couplings $b_{9/10/11}$, which turns out to be much
smaller than the  experimental error of
$\Delta\mu_{\Lambda\Sigma^0}^{\rm exp}=0.08$.
It is remarkable that the prediction for this physical quantity is so stable under
variation of $b_{9/10/11}$, although the fit values for the individual
couplings given in table~\ref{tab:magnmom} vary considerably.
While the heavy--baryon result is about two standard deviations off,
the relativistic one ($\mu_{\Lambda\Sigma^0}=1.606\pm0.008$~n.m.)
yields exactly the experimental result
(to be precise, experimentally only $|\mu_{\Lambda\Sigma^0}|$ is known).
We note furthermore that, while the values for the leading--order magnetic couplings
$b_6^{D/F}$ are fairly close in the two different schemes, those for the SU(3) breaking
couplings show no similarity at all. This clearly displays the fact that
$1/m$ corrections to the various loop diagrams are sizeable even beyond fourth order.


\subsection{Electric radii} 
\label{sec:elrad}

\noindent
The electric (or charge) radius of any baryon is, according to
eq.~(\ref{sachsdef}), given by
the sum of the Dirac radius and the so--called Foldy term,
\beq
\langle r_E^2 \rangle = \langle r_1^2 \rangle +\frac{3\,\kappa_B}{2\,m_B^2} ~.
\eeq
Phenomenologically  the Foldy term is hence well--known
\emph{for all ground state octet baryons}
from the experimental information on the magnetic moments.
What remains to be predicted from any kind of theory or model, as independent quantities,
are the Dirac radii.
We have therefore always replaced the chiral representation of the Foldy term
by the exact value given by experiment. This is legitimate at any chiral order,
as the difference is always subleading.
Doing otherwise would partly import the well--known problematic convergence properties
of the magnetic moments to the description of the electric
radii.\footnote{This can even entail, in our opinion, misleading conclusions:
the huge decuplet effects on the charge radii found in~\cite{puglia2}
in some cases stem from the Foldy term, hence have nothing to do with
intrinsically \emph{electric} properties of the baryons. It should
be noted that in a comparable SU(2) study, only minor effects due to
the $\Delta$ resonance were found~\cite{BFHM}.}

\begin{table}[ht] 
\begin{center} 
\renewcommand{\arraystretch}{1.2} 
\begin{tabular}{|c||c|c||c|c||c|} 
\hline
& \mc{HB} & \mc{IR} & \\
\hline 
& ${\cal O}(q^3)$ & ${\cal O}(q^4)$ &
  ${\cal O}(q^3)$ & ${\cal O}(q^4)$ & exp.\\ 
\hline 
$p$               &    0.717 &    0.717 &    0.717 &    0.717         &    0.717 \\
$n$               & $-$0.113 & $-$0.113 & $-$0.113 & $-$0.113         & $-$0.113 \\
$\Lambda$         &    0.14  &    0.00  &    0.05  &    0.11$\pm$0.02 & --- \\
$\Sigma^+$        &    0.59  &    0.72  &    0.63  &    0.60$\pm$0.02 & --- \\
$\Sigma^0$        & $-$0.14  & $-$0.08  & $-$0.05  & $-$0.03$\pm$0.01 & --- \\
$\Sigma^-$        &    0.87  &    0.88  &    0.72  &    0.67$\pm$0.03 & 0.60$\pm$0.08$\pm$0.08 \cite{selex}\\
                  &          &          &          &                  & 0.91$\pm$0.32$\pm$0.40 \cite{wa89}\\
$\Xi^0$           &    0.36  &    0.08  &    0.15  &    0.13$\pm$0.03 & --- \\
$\Xi^-$           &    0.67  &    0.75  &    0.56  &    0.49$\pm$0.05 & --- \\
$\Lambda\Sigma^0$ & $-$0.10  & $-$0.09  &    0.00  &    0.03$\pm$0.01 & --- \\
\hline
$d_{101}$         & $-$0.84  & $-$0.34  & $-$0.44  & $-$0.15          & \\
$d_{102}$         &    1.20  &    1.64  &    1.57  &    1.64          & \\
\hline
\end{tabular} 
\renewcommand{\arraystretch}{1.0} 
\caption{
Predictions for the electric radii $\langle r_E^2 \rangle$ [fm$^2$].
The various best--fit values for the pertinent counterterms
$d_{101}$, $d_{102}$ are given as well (in units of GeV$^{-2}$).
The experimental values for the proton and neutron are taken from
the dispersion theoretical studies~\cite{mmd,hmd}.
The errors for the relativistic fourth order predictions display
the uncertainty due to $\Delta b_6^{D/F}$.
The errors for the experimental $\Sigma^-$ radius
values refer to statistical (first) and systematic (second) errors.}
\label{tab:elrad} 
\end{center} 
\end{table}  

\medskip \noindent 
The chiral representations of the electric radii of the baryon octet,
both at third and fourth order,
involve exactly two low--energy constants, $d_{101}$ and $d_{102}$.
These can readily be fitted to the charge radii
of proton and neutron, such that all others can be predicted.
Table~\ref{tab:elrad} shows these predictions for third-- and fourth--order
calculations, both in the heavy--baryon and the infrared regularization
formalism.
The fourth column shows the predictions according to the fourth
order relativistic calculation
together with errors which reflect some theoretical uncertainty:
even with the Foldy term fixed,
some uncertainty inherited from the description of the magnetic moments remains.
Indeed, though kinematically suppressed
(i.e.\ of higher than fourth order in strict chiral power counting),
the loop corrections to the anomalous magnetic couplings, see diagram~(10) in
fig.~\ref{fig:diag}, contribute to the Dirac radius.
We employ the values for $b_6^{D/F}$ obtained from the best fit to the magnetic moments
at fourth order, $b_6^D=4.21$ and $b_6^F=1.64$, see table~\ref{tab:magnmom}.
In order to get an estimate of the uncertainty due to these LECs,
we again assume an error
of about half of the change when fitting them at third and fourth order,
thus assigning $\Delta b_6^{D/F}=0.5$. Table~\ref{tab:elrad} shows 
that \emph{this} uncertainty is (relatively) \emph{small}, as would be expected.
We would like to stress however that this error only indicates one particular effect.
An estimate of the complete uncertainty due to higher--order contributions is
hardly feasible (or, due to the appearance of new
unknown couplings, indeed impossible). 
Finally, the uncertainties due to the \emph{experimental} errors on the magnetic moments
entering the Foldy term are yet one order of magnitude smaller.

\medskip \noindent
With regard to convergence only, i.e.\ exclusively to the numerical changes
when going from third to fourth order,
the infrared regularization scheme yields overall considerable improvement
over the heavy--baryon results, especially for $\Lambda$, $\Sigma^+$, and $\Xi^0$.
This improvement can also be seen in the behavior of the fitted values
of $d_{101}$ and $d_{102}$, also given in table~\ref{tab:elrad},
which are more stable in the relativistic scheme.
What is more important though is that both the absolute values and the trends
within the two schemes  are entirely different for some hyperons.
E.g.\ for the $\Sigma^-$ radius,
the only hyperon radius on which experimental information exists,
the heavy--baryon values are very stable at 0.87--0.88~fm$^2$, but
deviate sizeably from the radius given by the SELEX collaboration~\cite{selex},
$\langle (r_E^{\Sigma^-})^2 \rangle = 0.60 \pm 0.08({\rm stat.})
\pm 0.08({\rm syst.})$~fm$^2$
(the pioneering WA89 measurement~\cite{wa89} is not precise enough
to favor any of the theoretical predictions).
In the relativistic scheme however,
the third--order value is already within this error range,
with the fourth--order one even closer to the central value.
Similarly for the other charged hyperons $\Sigma^+$ and $\Xi^-$,
the fourth--order corrections
\emph{increase} the third--order predictions in the heavy--baryon case,
but \emph{reduce} them in the relativistic scheme.
We also note that only the relativistic predictions show the hierarchy in
the size of the  electric radii expected from naive quark model considerations,
$\langle (r_E^p)^2 \rangle > \langle (r_E^{\Sigma^\pm})^2 \rangle >
\langle (r_E^{\Xi^-})^2 \rangle$.
The sizeable difference between the $\Sigma^+$ and $\Sigma^-$ radii at third order in the
heavy--baryon scheme is largely reduced in the relativistic results,
leaving only a 10\% effect at fourth order.
For the neutral hyperons,
all predictions are consistent as far as the signs of the radii are concerned
(with the exception of the radius of the $\Lambda$--$\Sigma^0$ transition form factor),
yielding positive radii for the $\Lambda$ and $\Xi^0$ hyperons, and a negative radius
for the $\Sigma^0$.
Quantitatively, the relativistic fourth--order calculation
predicts radii of a size very similar to that of the neutron for $\Lambda$ and $\Xi^0$,
and a $\Sigma^0$ radius of about another factor of 3 smaller.

\medskip \noindent
To conclude this section, we would like to comment on the issue of SU(3) breaking
in the electric radii. In contrast to the calculation of the magnetic
moments to fourth order,
the electric radii to this order do not contain any operators which break SU(3)
at tree level (compare the LECs $\alpha_{1-4},\,\beta_1$ in eq.~(\ref{Lagr4}) in the
magnetic sector), therefore all SU(3) breaking effects are created `dynamically'
via mass splittings in the loop diagrams. The leading (and dominating) effect
is the pion--kaon mass difference. We remind the reader that SU(3) symmetry would
give the hyperon radii in terms of the proton/neutron radii according to
\beqa
      \langle ( r_E^{\Sigma^+} )^2 \rangle \!\!&=&\!\!      \langle ( r_E^{p}               )^2 \rangle ~, \qquad
      \langle ( r_E^{\Sigma^-} )^2 \rangle   \, = \,        \langle ( r_E^{\Xi^-}           )^2 \rangle
\,=\, \langle ( r_E^{p}        )^2 \rangle      +           \langle ( r_E^{n}               )^2 \rangle ~, \no\\
  2\, \langle ( r_E^{\Lambda}  )^2 \rangle \!\!&=&\!\! -2\, \langle ( r_E^{\Sigma^0}        )^2 \rangle
\,=\, \langle ( r_E^{\Xi^0}    )^2 \rangle
\,=\,                    - \frac{2}{\sqrt{3}}\,             \langle ( r_E^{\Lambda\Sigma^0} )^2 \rangle
\,=\, \langle ( r_E^{n}        )^2 \rangle ~.
\eeqa
It is obvious that SU(3) breaking due to the large kaon mass changes this pattern
considerably: exact SU(3) predicts
$\langle ( r_E^{\Sigma^+} )^2 \rangle > \langle ( r_E^{\Sigma^-} )^2 \rangle$
which is reversed in all four ChPT results presented above. In addition, it
changes the sign of all charge radii for neutral hyperons.
The size of the additional SU(3) breaking due to the mass splitting
in the baryon octet, see diagrams~(5$^*$)--(7$^*$) in fig.~\ref{fig:diag}, 
can be seen by comparing the predictions at third and fourth
order in the relativistic framework, the change being mainly due to these
mass differences
(in fact, in terms of strict chiral power counting, this is the \emph{only} new
effect at fourth order apart from $1/m$ corrections to third order loops).
The net effect is a further moderate reduction of the charged hyperon radii.
Among the neutral particles, only the $\Lambda$ radius shows a sizeable correction.
Going to fifth order, one would obtain a large amount of additional
SU(3) breaking effects (e.g.\ in the meson--baryon coupling constants or
`explicit' breaking in the charge radii via contact terms) none of which is
quantifiable to sufficient accuracy. 
We thus regard our (fourth--order relativistic)
predictions as the best one is ever likely to achieve in any ChPT approach.


\subsection{Magnetic radii} 
\label{sec:magnrad} 

As the electric radii, the magnetic radii at fourth order include two parameters
(labeled $\eta_{1/2}$ in eq.~(\ref{Lagr4})) which can be fitted to the respective
proton and neutron data (taken here from the dispersive analysis~\cite{mmd,hmd})
in order to yield predictions for the hyperons.
However, in contrast to the electric case, here loops proportional to other
rather poorly known LECs ($b_{9/10/11}$) contribute significantly. 
As there are no further measurements of magnetic radii which would allow to 
fit these, the most transparent
thing to do is to indicate the uncertainty following from this poor knowledge.
Table~\ref{tab:magnrad} shows these uncertainties, based on $\Delta b_{9/10/11}=0.25$~GeV$^{-1}$
(which are assumed to be uncorrelated errors).
Again, this is only one particular effect and does not reflect
all possible uncertainties due to higher--order contributions. In
addition,  errors on these couplings were also only roughly estimated, such that they could
even be larger.

\begin{table}[ht] 
\begin{center} 
\renewcommand{\arraystretch}{1.2} 
\begin{tabular}{|c||c||c||c|} 
\hline 
& ${\cal O}(q^4)$ ${\rm HB}$ & ${\cal O}(q^4)$ ${\rm IR}$ & exp.\\ 
\hline 
$p$               &    0.699         & 0.699         & 0.699 \\
$n$               &    0.790         & 0.790         & 0.790 \\
$\Lambda$         &    0.30$\pm$0.11 & 0.48$\pm$0.09 & --- \\
$\Sigma^+$        &    0.74$\pm$0.06 & 0.80$\pm$0.05 & --- \\
$\Sigma^0$        &    0.20$\pm$0.10 & 0.45$\pm$0.08 & --- \\
$\Sigma^-$        &    1.33$\pm$0.16 & 1.20$\pm$0.13 & --- \\
$\Xi^0$           &    0.44$\pm$0.15 & 0.61$\pm$0.12 & --- \\
$\Xi^-$           &    0.44$\pm$0.20 & 0.50$\pm$0.16 & --- \\
$\Lambda\Sigma^0$ &    0.60$\pm$0.10 & 0.72$\pm$0.10 & --- \\
\hline
$\eta_1$          &    0.26$\pm$0.10 & 0.69$\pm$0.10 & \\
$\eta_2$          & $-$0.02$\pm$0.21 & 0.72$\pm$0.21 & \\
\hline
\end{tabular} 
\renewcommand{\arraystretch}{1.0} 
\caption{
Predictions for the magnetic radii $\langle r_M^2 \rangle$ [fm$^2$].
The various best--fit values for the pertinent counterterms
$\eta_1$, $\eta_2$ are given as well (in units of GeV$^{-3}$).
The errors display the uncertainty due to $\Delta b_{9/10/11}$.} 
\label{tab:magnrad} 
\end{center} 
\end{table}  

\medskip \noindent
Nevertheless, we consider the errors in table~\ref{tab:magnrad}
indicative enough to state that the magnetic radii can be predicted only to much
lower accuracy than the electric ones. Given the sizeable uncertainties,
there is no significant discrepancy between the heavy--baryon and the relativistic
predictions (with the exception of the $\Sigma^0$).
The general trend is an increase of most magnetic radii in the relativistic
scheme compared to the heavy--baryon results, where the central values are surprisingly small
for some hyperons ($\Lambda$, $\Sigma^0$).
The values for the LECs $\eta_{1/2}$, however, are completely different.
This indicates large SU(3) symmetric loop contributions that have to be canceled by
these counterterms. This does not come as a surprise
if one remembers what is known about loop contributions to the magnetic moments.
Note that with regard to the values for the SU(3) breaking terms in the magnetic moments
given in section~\ref{sec:magnmom}, \emph{none} of the fourth--order couplings
can be fixed in agreement with \emph{both} schemes.
SU(3) breaking is large, as a completely SU(3) symmetric treatment of the
magnetic form factors (moments \emph{and} radii) would imply
\beqa
      \langle ( r_M^{\Sigma^+} )^2 \rangle \!\!&=&\!\!  \langle ( r_M^{p}     )^2 \rangle ~, \no\\
      \langle ( r_M^{\Sigma^-} )^2 \rangle \!\!&=&\!\!  \langle ( r_M^{\Xi^-} )^2 \rangle
\,=\, \frac{\mu_p \langle ( r_M^{p} )^2 \rangle + \mu_n \langle ( r_M^{n}     )^2 \rangle}{\mu_p+\mu_n} ~, \no\\
      \langle ( r_M^{\Lambda}  )^2 \rangle \!\!&=&\!\!  \langle ( r_M^{\Sigma^0} )^2 \rangle
\,=\, \langle ( r_M^{\Lambda\Sigma^0} )^2 \rangle \,=\, \langle ( r_M^{n}        )^2 \rangle ~.
\eeqa
In contrast to these, the values in table~\ref{tab:magnrad} show no clear pattern.
The magnetic radius of the $\Sigma^-$ is remarkable in being much larger
than all other radii, electric or magnetic. Such an effect, albeit less
dramatic, is also found in some lattice studies, see e.g.~\cite{LWD}.
(This study also predicts relatively small magnetic radii
for $\Lambda$ and $\Xi^-$ as we do, though not for the $\Sigma^0$.)
As in the case of the electric radii,
improvement of these predictions in the framework of ChPT is hardly
feasible as higher order corrections would include numerous
unknown SU(3) breaking effects.

\subsection{\boldmath{$Q^2$}--dependence of the form factors} 
\label{sec:q2dep} 

\noindent
In~\cite{irff} it was shown that the complete relativistic chiral
one--loop representation
fails to describe the $Q^2$--dependence of the `large' form factors (i.e.\ those not
vanishing at $Q^2=0$) already at rather low $Q^2$. As a remedy, it was demonstrated that
the inclusion of dynamical vector mesons, used in an antisymmetric tensor representation
and coupled to nucleons/pions/photons in a chirally invariant fashion, yields a very
good description of all electromagnetic nucleon form factors up to about 0.4~GeV$^2$.
Certainly, in order to obtain reasonable predictions for the $Q^2$--dependence
of the hyperon form factors, one has to proceed likewise here.
The necessary formalism and the definitions of all pertinent couplings
are presented in great detail in~\cite{irff}. Of course one has
to assume SU(3) symmetric vector meson couplings to the baryons, which are
then fully determined by the values for the vector meson nucleon couplings
as given in~\cite{mmd}. As these transform in the same way as the contact terms,
replacing part of the latter by explicit vector meson contributions on tree level affects
in no way the predictions for the hyperon radii.

\begin{figure}[htb]
\centerline{
\epsfysize=4.4in
\epsffile{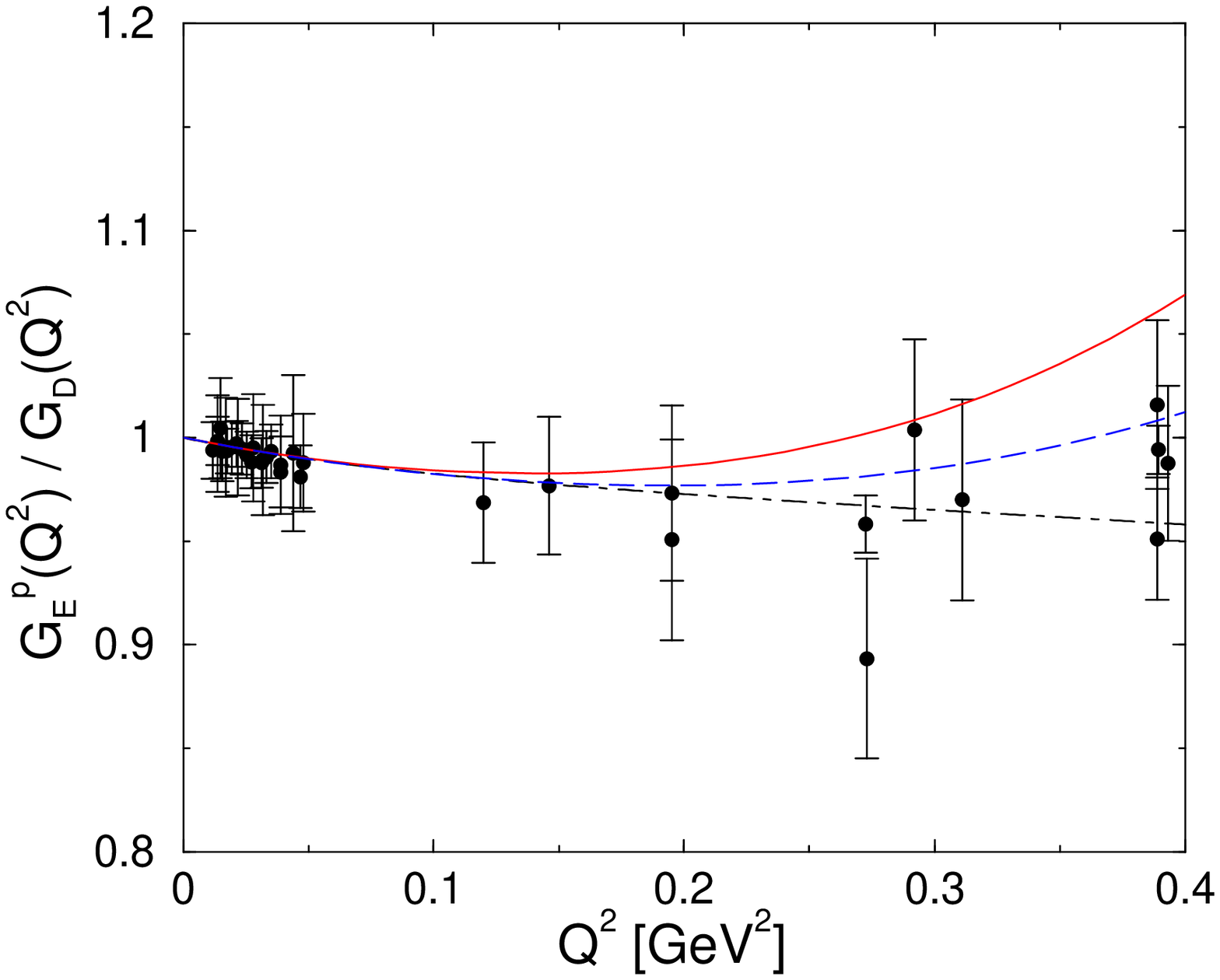}
}
\vskip 2.5cm
\caption{The proton electric form factor in SU(3) (red/full curve) 
and SU(2) (blue/dashed curve) relativistic baryon chiral
perturbation theory including vector mesons, divided by the dipole form factor.
For comparison, we show the dispersion theoretical result (black dot--dashed curve)
and the world data available in this energy range.
\label{fig:GEp}
}
\end{figure}

\begin{figure}[htb]
\centerline{
\epsfysize=4.4in
\epsffile{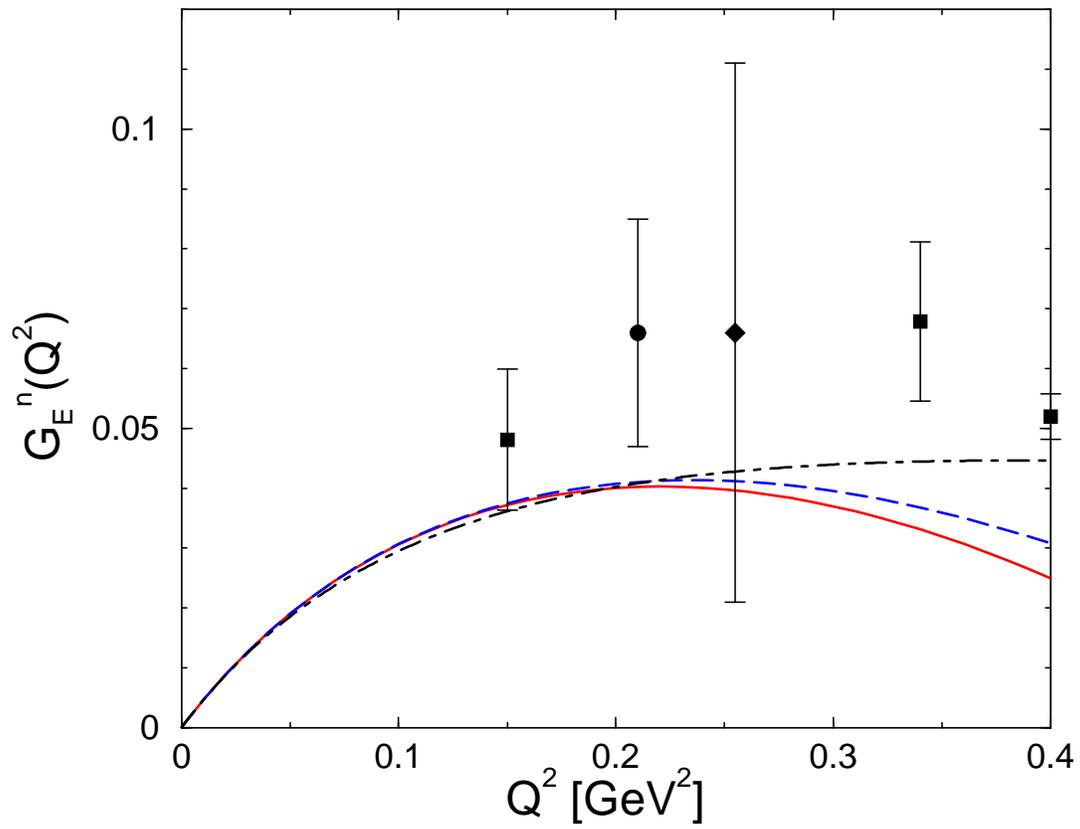}
}
\vskip 2.5cm
\caption{The neutron electric form factor in SU(3) (red/full curve) 
and SU(2) (blue/dashed curve) relativistic baryon chiral
perturbation theory including vector mesons.
Also given is the result of  the dispersion
theoretical analysis (black dot--dashed curve).
We only show the more recent data as given in 
\protect\cite{newnff}.
\label{fig:GEn}
}
\end{figure}

\medskip \noindent
The fourth order results for the electric form factors of proton and neutron,
including vector meson effects, are shown in figs.~\ref{fig:GEp} and \ref{fig:GEn},
respectively, the former divided by the dipole form factor. For comparison we
show the equivalent SU(2) results given in~\cite{irff} and the dispersion theoretical
fits from~\cite{mmd,hmd}.
In both cases the difference between the SU(2) and SU(3) descriptions is very small,
though the SU(3) one is slightly worse. From such small differences one concludes
that not much room is left for a strangeness contribution via kaon loops, as
simple meson cloud models seem to indicate. Strangeness as hidden in the $\phi$--meson
component or from higher mass states encoded in the value of the LECs $d_{101/102}$
cannot be separated from the analysis presented here. To completely disentangle
the strangeness contribution to a given form factor, a full flavor decomposition
is necessary. For that, one also has to calculate the singlet form factors since the
electromagnetic current is only sensitive to the triplet and octet current components.

\begin{figure}[htb]
\centerline{
\epsfysize=4.4in
\epsffile{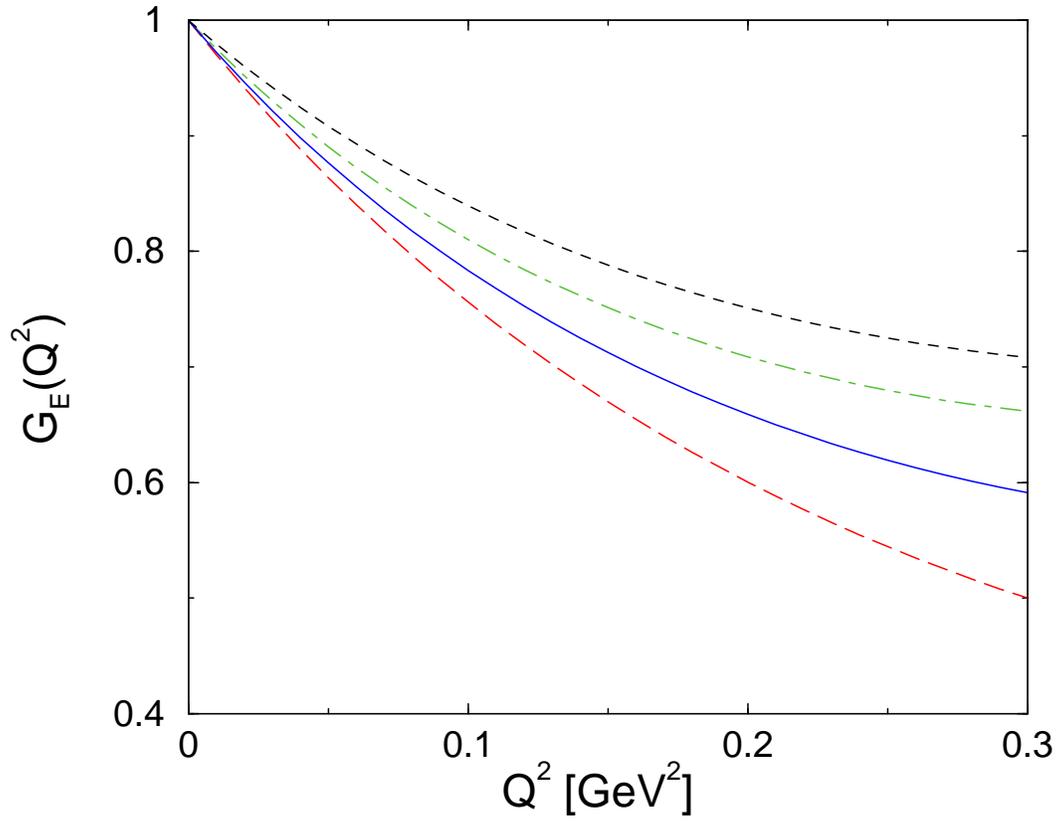}
}
\vskip 2.5cm
\caption{
Predictions for
the electric form factors of the charged hyperons $\Sigma^+$ (green/dot--dashed line),
$\Sigma^-$ (blue/full line), and $\Xi^-$ (black/dashed line).
For the latter two, the absolute value of $G_E(Q^2)$ is shown.
For comparison, we also show the proton electric form factor (red/long--dashed line).
\label{fig:GEcharged}
}
\end{figure}

\begin{figure}[htb]
\centerline{
\epsfysize=4.4in
\epsffile{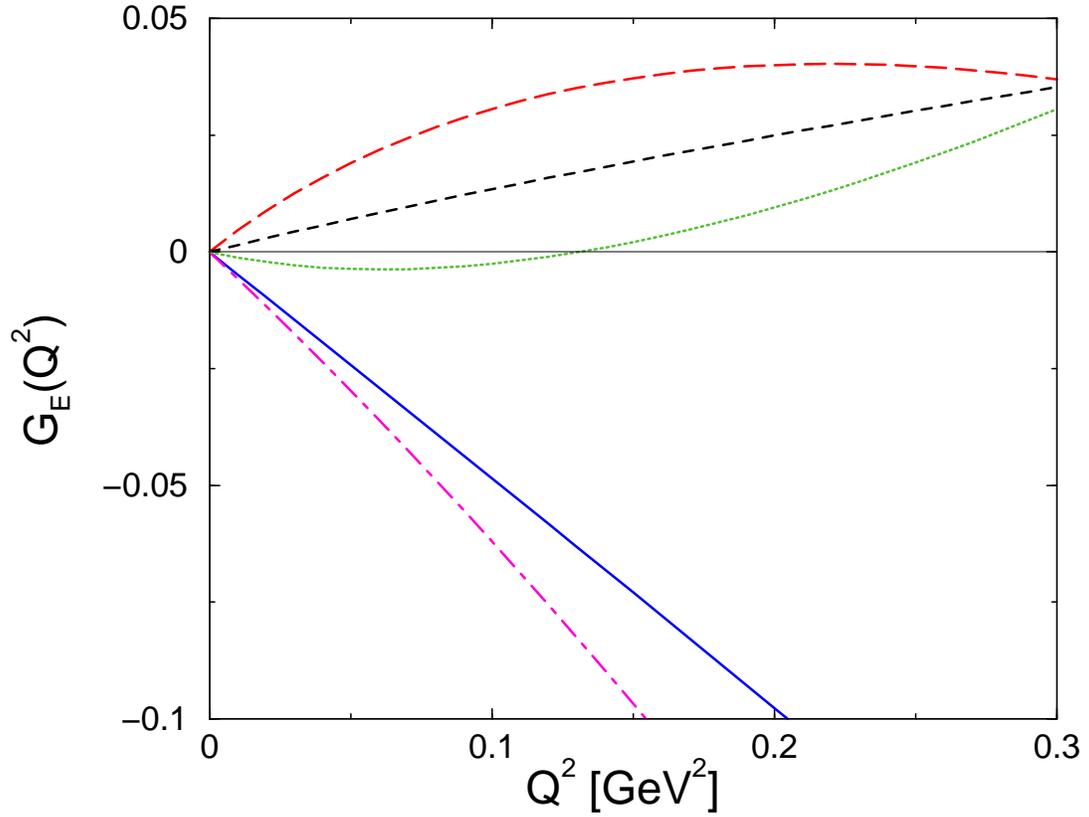}
}
\vskip 2.5cm
\caption{
Predictions for
the electric form factors of the neutral hyperons $\Lambda$ (blue/full line),
$\Sigma^0$ (black/dashed line), $\Xi^0$ (pink/dot--dashed line),
as well as the $\Lambda$--$\Sigma^0$ transition form factor (green/dotted line).
For comparison, we also show the neutron electric form factor (red/long--dashed line).
\label{fig:GEneutral}
}
\end{figure}

\medskip \noindent
Fig.~\ref{fig:GEcharged} shows the electric form factors of all charged hyperons
(those of the negative ones with sign reversed). They all show a $Q^2$--dependence
qualitatively similar to that of the proton charge form factor,
the quantitative difference being due to their different
sizes as determined by the radii.
In all cases, the vector meson effects contribute a large part of the curvature which
is far too small for the proton in a purely chiral representation.
The neutral hyperon form factors are shown in
fig.~\ref{fig:GEneutral}
in comparison to the neutron electric form factor.
Similar to what was found for the latter in~\cite{irff},
the vector meson effects largely cancel for all charge form
factors of neutral hyperons.
Apart from the neutron, only the $\Lambda$--$\Sigma^0$ transition form factor shows
significant curvature, the ones for $\Lambda$, $\Sigma^0$, and $\Xi^0$ are dominated
by the radius term and display a nearly linear behavior.

\medskip \noindent
For the magnetic form factors, a problem occurs when trying to transfer
the procedure to include vector mesons exactly from the SU(2) to the SU(3) case.
In the former, also loop corrections to the tree level diagrams including
vector meson exchange were calculated, in strict analogy to loop corrections
to the leading order magnetic couplings at fourth order, see diagrams (10)--(12)
in fig.~\ref{fig:diag}. However, in contrast to the SU(2) case, these
loop corrections are \emph{large} in SU(3), such that there is no reason
to identify the bare couplings with the ones determined in a dispersive analysis.
What is more, in contrast to the electric form factors, such loops would lead
to additional SU(3) breaking effects in the magnetic radii, yielding largely
different values compared to a strictly chiral analysis. To avoid
these problems, one should use vector mesons on tree level only in an SU(3) analysis,
producing predictions for the magnetic form factors in agreement with the chiral
predictions for the magnetic radii.

\begin{figure}[htb]
\centerline{
\epsfysize=4.4in
\epsffile{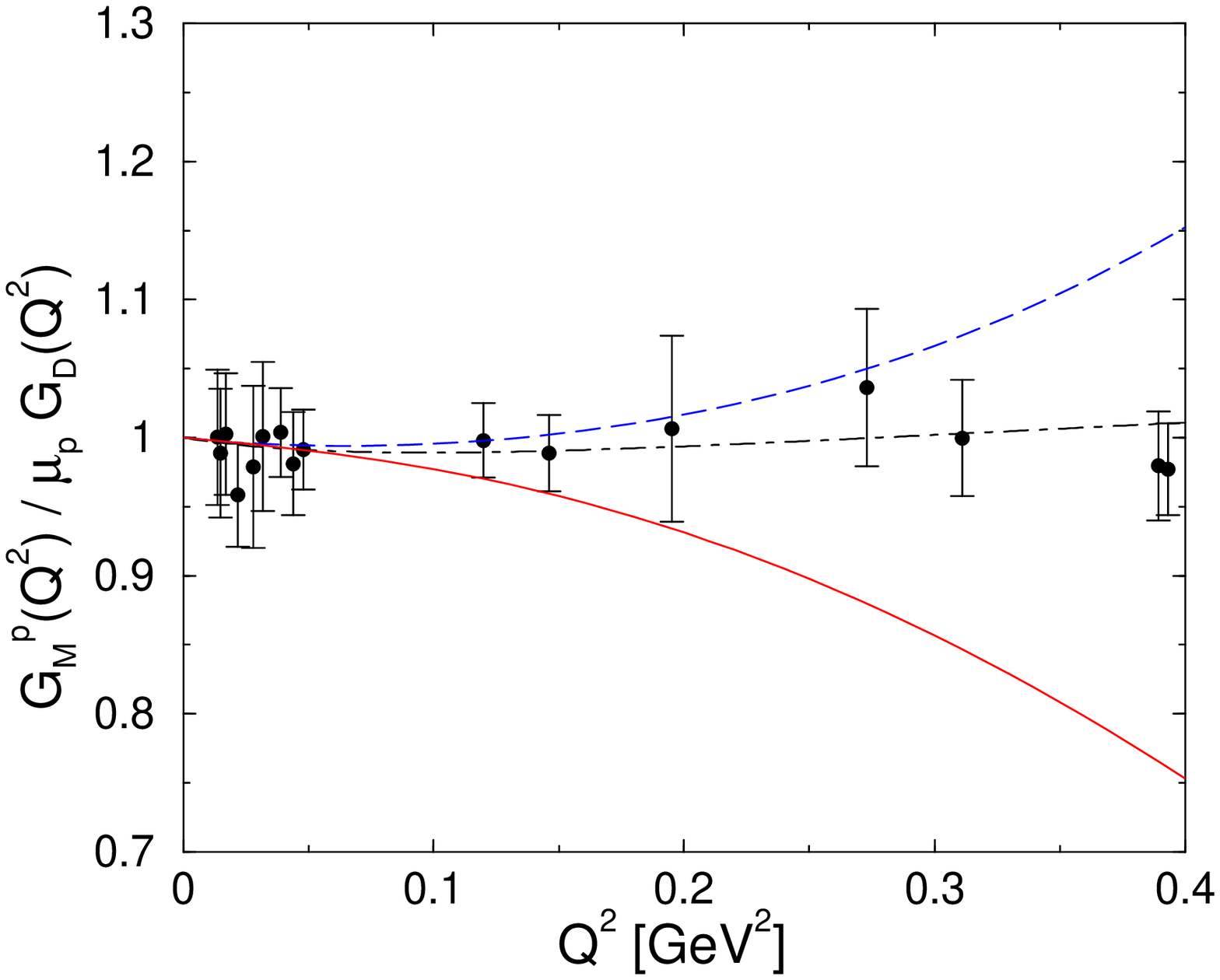}
}
\vskip 2.5cm
\caption{The proton magnetic form factor in SU(3) (red/full curve) 
and SU(2) (blue/dashed curve) relativistic baryon chiral
perturbation theory including vector mesons, divided by the dipole form factor.
For comparison, we show the dispersion theoretical result (black dot--dashed curve)
and the world data available in this energy range.
\label{fig:GMp}
}
\end{figure}

\begin{figure}[htb]
\centerline{
\epsfysize=4.4in
\epsffile{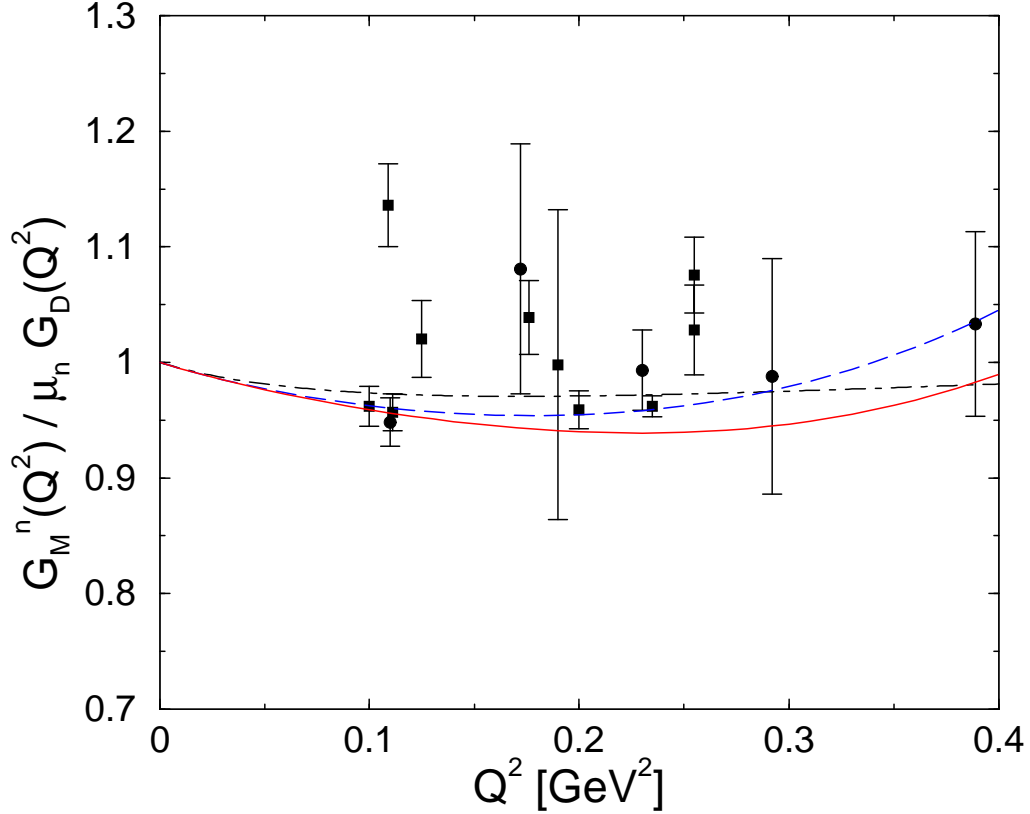}
}
\vskip 2.1cm
\caption{The neutron magnetic form factor in SU(3) (red/full curve) 
and SU(2) (blue/dashed curve) relativistic baryon chiral
perturbation theory including vector mesons, divided by the dipole form factor.
For comparison, we show
the dispersion theoretical result (black dot--dashed curve)
and the world data available in this energy range, where the data points
denoted by squares (instead of circles) refer to the more recent
measurements~\protect\cite{gmnnew}.
The older data can be traced back from~\protect\cite{mmd}.
\label{fig:GMn}
}
\end{figure}

\medskip \noindent
We only show the magnetic form factors of proton and neutron
divided by the dipole form factor,
see figs.~\ref{fig:GMp},~\ref{fig:GMn}, again compared to the
SU(2) analysis in~\cite{irff}.
The proton magnetic form factor proves to be worse above 0.2~GeV$^2$
in the SU(3) case, much closer to the third order curve given in~\cite{irff}
(which also does not include loop corrections to vector meson couplings).
The curvature in the SU(3) calculation is slightly too small to meet the data
above 0.2~GeV$^2$.
For the neutron, however, there is hardly any difference between the two descriptions.
The general observation for the hyperons (not displayed here) is that
the neutral ones, like the neutron, tend to have stronger curvature, while the magnetic
form factors of the charged hyperons are closer to a purely chiral description
dominated by the radius term, and probably suffer from a similar deficit as the
proton magnetic form factor.

\section{Summary} 
\label{sec:sum} 
\def\theequation{\arabic{section}.\arabic{equation}} 
\setcounter{equation}{0}

We have studied the electromagnetic form factors of the baryon octet
in a manifestly Lorentz invariant form of baryon chiral perturbation
theory to one--loop (fourth) order employing the
so--called infrared regularization of loop graphs.
The pertinent results of our investigation can be summarized as follows:
\begin{itemize} 
\item[(1)] We have argued that the chiral expansion of the magnetic
  moments in the relativistic scheme is ambiguous at third order,
  such that no clear statement can be made whether or not
  convergence is improved in comparison to the heavy--baryon scheme.
  At fourth order, due to
  the presence of seven low--energy constants, one can
  only predict the $\Lambda$--$\Sigma^0$ transition moment,
  $\mu_{\Lambda\Sigma^0} = 1.61 \pm 0.01$~n.m., in stunning agreement with
  the empirical value.
\item[(2)] To fourth order, only two LECs affect the electric radii.
  These can be fixed from the measured neutron and proton radii. Always
  using the empirical magnetic moments in the Foldy term, we have shown
  that the fourth order corrections to the electric radii are astonishingly
  small, and so are the resulting uncertainties.
  The prediction for the $\Sigma^-$ radius agrees with the recent
  result from the SELEX collaboration~\cite{selex}. The pion--kaon mass
  difference leads to sizeable deviations from flavor SU(3) symmetry.
\item[(3)] The magnetic radii cannot be predicted so
  precisely. Again, one finds large SU(3) breaking due to loop
  corrections. In particular, the
  magnetic radius of the $\Sigma^-$ is largest.
\item[(4)] For the electric form factors of the charged particles,
  the pure chiral representation
  provides too little curvature. With vector mesons included as 
  in~\cite{irff}, the $Q^2$--dependence of various charged form factors
  is given up to virtualities of $Q^2 = 0.3$~GeV$^2$,
  see figs.~\ref{fig:GEp}, \ref{fig:GEcharged}. For the neutral
  particles we find in general a large cancellation of these vector
  meson contributions, and the resulting form factors for the neutral hyperons
  display less curvature than the neutron one,
  see figs.~\ref{fig:GEn}, \ref{fig:GEneutral}.
  We do not observe any sizeable effects in the
  electric proton and neutron form factors when going from SU(2) to
  SU(3). Again, the corresponding magnetic form factors cannot be
  predicted so precisely.
\end{itemize}

\medskip\noindent
In the future, it will be of interest to also calculate the singlet electromagnetic
currents. This would allow one to perform a flavor decomposition of the various form factors
and in particular to reanalyze the so--called strange form factors of the nucleon, which
are currently of great interest and have been studied in the heavy--baryon limit only~\cite{ito,hms,hkm}.

\section*{Acknowledgements}
We would like to thank Nadia Fettes for a careful reading of the manuscript.


\bigskip


$\,$
\vskip 1.0cm

\noindent {\Large {\bf Figures}}

\vskip 3.0cm

\pagebreak

$\,$
\vskip 3cm

\pagebreak

$\,$
\vskip 3cm

\pagebreak

$\,$
\vskip 3cm

\pagebreak

$\,$
\vskip 3cm

\pagebreak

$\,$
\vskip 3cm

\pagebreak

$\,$
\vskip 2.5cm

\end{document}